\documentclass[referee]{aa}
\usepackage{graphicx}
\usepackage{txfonts}

\usepackage{subfig}
\usepackage{float}
\usepackage{comment}
\usepackage{caption}
\usepackage{bm}
\usepackage{amsfonts}
\usepackage{amssymb}
\usepackage{array}
\usepackage{natbib}
\usepackage{tikz}
\usepackage{booktabs}
\newcommand{\argmin}{\mathop{\mathrm{arg\:min}}}

\usepackage{url}

\usepackage{color}
\usepackage{multirow}

\usepackage{multicol}
\usepackage{colortbl, dcolumn}
\usepackage{xcolor}

\newcommand{\R}{\mathbb{R}}

\newcommand{\abs}[1]{\left\vert#1\right\vert}

\usepackage[ruled,vlined]{algorithm2e}

\begin{document}

\title{Forward-fitting STIX visibilities}

\author{Anna Volpara\inst{1} \and Paolo Massa\inst{1} \and Emma Perracchione\inst{2}  \and Andrea Francesco Battaglia\inst{4,5} \and Sara Garbarino\inst{1} \and Federico Benvenuto\inst{1} \and Anna Maria Massone\inst{1,3} \and S\"am Krucker\inst{4,6} \and Michele Piana\inst{1,3}}
\institute{MIDA, Dipartimento di Matematica,  Universit\`{a} degli Studi di Genova, Via   Dodecaneso 35, 16146 Genova, Italy
\and
Dipartimento di Scienze Matematiche \lq \lq Giuseppe Luigi Lagrange\rq \rq, Politecnico di Torino, Corso Duca degli Abruzzi, 24, 10129, Torino, Italy
\and
CNR-SPIN, Via Dodecaneso 33, 16146 Genova, Italy
\and
University of Applied Sciences and Arts Northwestern Switzerland, Bahnhofstrasse 6, 5210 Windisch, Switzerland
\and
ETH Z\"urich, R\"amistrasse 101, 8092 Z\"urich, Switzerland
\and
Space Sciences Laboratory, University of California, 7 Gauss Way, 94720 Berkeley, USA\\
\\
\email{volpara@dima.unige.it},
\email{massa.p@dima.unige.it},
\email{emma.perracchione@polito.it},
\email{andrebat@ethz.ch},
\email{benvenuto@dima.unige.it},
\email{massone@dima.unige.it},
\email{krucker@berkeley.edu},
\email{piana@dima.unige.it},
}


\abstract
{}
{To determine to what extent the problem of forward fitting visibilities measured by the Spectrometer/Telescope Imaging X-rays (STIX) on-board Solar Orbiter is more challenging with respect to the same problem in the case of previous hard X-ray solar imaging missions; to identify an effective optimization scheme for parametric imaging for STIX.}
{This paper introduces a Particle Swarm Optimization (PSO) algorithm for forward fitting STIX visibilities and compares its effectiveness with respect to the standard simplex-based optimization algorithm used so far for the analysis of visibilities measured by the Reuven Ramaty High Energy Solar Spectroscopic Imager (RHESSI). This comparison is made by considering experimental visibilities measured by both RHESSI and STIX, and synthetic visibilities generated by accounting for the STIX signal formation model.}
{We found out that the parametric imaging approach based on PSO is as reliable as the one based on the simplex method in the case of RHESSI visibilities. However, PSO is significantly more robust when applied to STIX simulated and experimental visibilities.}
{Standard deterministic optimization is not effective enough for forward-fitting the few visibilities sampled by STIX in the angular frequency plane. Therefore a more sofisticated optimization scheme must be introduced for parametric imaging in the case of the Solar Orbiter X-ray telescope. The forward-fitting routine based on PSO we introduced in this paper proved to be significantly robust and reliable, and could be considered as an effective candidate tool for parametric imaging in the STIX context.}


\titlerunning{Forward-fitting STIX visibilities}
\authorrunning{Volpara et al}

\maketitle

\date{\today}

\section{Introduction}

Fourier imaging in astronomy has been first introduced in radio astrophysics. Indeed, interferometric arrays of on-Earth radio telescopes record sets of Fourier components of the incoming photon flux, named visibilities. The first use of visibilities in space instruments probably dates back to around forty years ago during the Yohkoh mission \citep{kosugi1991hard}. However, the first space instrument utilizing visibility-based imaging in an extensive and systematic way has been designed by G. Hurford at the beginning of this century, with the NASA RHESSI \citep{2002SoPh..210....3L}
 mission for the hard X-ray observation of solar flares \citep{2002SoPh}. Nowadays, visibilities are the native form of measurements for the Spectrometer/Telescope for Imaging X-rays (STIX) \citep{krucker2020spectrometer}, which is one of the remote sensing instruments on-board the ESA Solar Orbiter mission. 

There are crucial differences between the processing of radio and hard X-ray visibilities. On the one hand, the dishes of radio telescopes are typically very large and hence collect very large numbers of photons (of the order of several thousands) characterized by very small energy per photon. On the other hand, RHESSI and STIX observe a much smaller number of photons characterized by much higher energies. For RHESSI, the Nyquist theorem and the requirement for an adequate sampling of a modulation cycle of its Rotating Modulation Collimators \citep{pianabook} imply that the number of independent visibilities available for imaging purposes is around 300 (however, statistically significant visibilities are often fewer, typically ranging between a few dozens and one hundred). For STIX, things are even worse: the sampled frequency points in this case are determined by the number and geometry of the sub-collimators, which are just 30.

Physical and geometrical constraints have significant impacts on the way visibilities are processed for image reconstruction. In the case of radio astronomy, interpolation in the visibility space and application of inverse Fourier transform are sufficient to obtain reliable images of the radio source \citep{2021A&A...646A..58A}. In the RHESSI and even more in the STIX frameworks, more sophisticated mathematics based on regularization theory for ill-posed problems is needed to reduce the ambiguity induced by such a sparse sampling of the data space \citep{2022arXiv220209334M,2021ApJ...919..133P,2020mem_ge,2009ApJ...703.2004M}.

Among visibility-based hard X-ray imaging methods, forward-fitting algorithms aim at estimating the input parameters of pre-defined shapes of the emitting sources by means of optimization procedures. The idea of these algorithms is to model an emitting source by means of a parameterized function (typically obtained by modifications and/or replications of a two-dimensional Gaussian function) and then to compute the parameter values that minimize the sum of the squared distance between the observed and predicted visibilities. 
The crucial tool for implementing these approaches is the optimization scheme applied to realize such minimization. 
Almost all RHESSI studies relied on simplex Nelder-Mead optimization, also known as AMOEBA Search \citep{press2007numerical,2002SoPh..210..193A,2002SoPh}, with results that are always reliable and robust \citep{2008ApJ...673..576X,2011ApJ...730L..22K,2012A&A...543A..53G,2012ApJ...755...32G,2013ApJ...766...28G,2018ApJ...867...82D}. The only exception to this simplex-based approach is probably represented by a study of the February 20 2002 event where visibility forward-fitting is realized by means of a Sequential Monte Carlo sampler \citep{2018ApJ...862...68S}.

The rationale of the present study was to assess whether this same optimization scheme is reliable as well in the case of STIX visibilities. The result of our analysis is that, when the number of available visibilities is so small as in the case for the ESA space telescope, there are significant experimental conditions where the use of AMOEBA for optimization provides unreliable parameter estimation. Therefore, here we describe the performances of a different minimization approach, based on a biology-inspired strategy \citep{wahde2008biologically,kennedy1995particle,liu2011survey} that has been used for fitting STIX visibility amplitudes when fully calibrated visibilities were not yet available \citep{2021A&A...656A..25M}. In the present paper this strategy is applied for the first time to fully calibrated visibilities and its performances are compared with the ones of simplex optimization in the case of several data sets made of both synthetic and experimental STIX visibilities. 

The plan of the paper is as follows. Section 2 provides a simple formalism illustrating the forward-fit problem for STIX. Section 3 introduces the biology-inspired optimization algorithm. Section 4 contains the results of the analysis of both synthetic and observed STIX visibilities. Our comments and conclusions are offered in Section 5.

\section{The STIX forward-fit problem}

\begin{figure*}[h]
\begin{center}

\tikzset{every picture/.style={line width=0.70pt}} 

\begin{tikzpicture}[x=0.70pt,y=0.70pt,yscale=-1,xscale=1]

\draw    (38,180) -- (37.2,19.8) ;
\draw [shift={(37.19,16.8)}, rotate = 89.71] [fill={rgb, 255:red, 0; green, 0; blue, 0 }  ][line width=0.08]  [draw opacity=0] (7.14,-3.43) -- (0,0) -- (7.14,3.43) -- cycle    ;
\draw    (20,163) -- (192,163) ;
\draw [shift={(195,163)}, rotate = 180] [fill={rgb, 255:red, 0; green, 0; blue, 0 }  ][line width=0.08]  [draw opacity=0] (7.14,-3.43) -- (0,0) -- (7.14,3.43) -- cycle    ;
\draw  [color={rgb, 255:red, 74; green, 144; blue, 226 }  ,draw opacity=1 ][line width=1.5]  (59,95) .. controls (59,78.43) and (72.43,65) .. (89,65) .. controls (105.57,65) and (119,78.43) .. (119,95) .. controls (119,111.57) and (105.57,125) .. (89,125) .. controls (72.43,125) and (59,111.57) .. (59,95) -- cycle ;
\draw  [dash pattern={on 0.84pt off 2.51pt}]  (89,95) -- (90,162) ;
\draw  [dash pattern={on 0.84pt off 2.51pt}]  (89,96) -- (36,96) ;
\draw [color={rgb, 255:red, 65; green, 117; blue, 5 }  ,draw opacity=1 ]   (67,73) -- (110,117) ;
\draw    (329,180) -- (328.2,19.8) ;
\draw [shift={(328.19,16.8)}, rotate = 89.71] [fill={rgb, 255:red, 0; green, 0; blue, 0 }  ][line width=0.08]  [draw opacity=0] (7.14,-3.43) -- (0,0) -- (7.14,3.43) -- cycle    ;
\draw    (301,163) -- (489,163) ;
\draw [shift={(492,163)}, rotate = 180] [fill={rgb, 255:red, 0; green, 0; blue, 0 }  ][line width=0.08]  [draw opacity=0] (7.14,-3.43) -- (0,0) -- (7.14,3.43) -- cycle    ;
\draw  [dash pattern={on 0.84pt off 2.51pt}]  (380,95) -- (381,162) ;
\draw  [dash pattern={on 0.84pt off 2.51pt}]  (425,95) -- (327,96) ;
\draw  [color={rgb, 255:red, 74; green, 144; blue, 226 }  ,draw opacity=1 ][line width=1.5]  (336.83,114.02) .. controls (330.27,98.31) and (344.28,77.51) .. (368.12,67.55) .. controls (391.97,57.6) and (416.61,62.27) .. (423.17,77.98) .. controls (429.73,93.69) and (415.72,114.49) .. (391.88,124.45) .. controls (368.03,134.4) and (343.39,129.73) .. (336.83,114.02) -- cycle ;
\draw  [draw opacity=0] (400.29,87.3) .. controls (403.65,89.7) and (405.84,92.65) .. (406.59,95.92) -- (368.54,102.43) -- cycle ; \draw   (400.29,87.3) .. controls (403.65,89.7) and (405.84,92.65) .. (406.59,95.92) ;  
\draw [color={rgb, 255:red, 65; green, 117; blue, 5 }  ,draw opacity=1 ]   (368.12,67.55) -- (391.88,124.45) ;
\draw [color={rgb, 255:red, 65; green, 117; blue, 5 }  ,draw opacity=1 ]   (336.83,114.02) -- (423.17,77.98) ;
\draw    (394,425) -- (393.02,232) ;
\draw [shift={(393,229)}, rotate = 89.71] [fill={rgb, 255:red, 0; green, 0; blue, 0 }  ][line width=0.08]  [draw opacity=0] (7.14,-3.43) -- (0,0) -- (7.14,3.43) -- cycle    ;
\draw    (308,406) -- (506,405.01) ;
\draw [shift={(509,405)}, rotate = 179.71] [fill={rgb, 255:red, 0; green, 0; blue, 0 }  ][line width=0.08]  [draw opacity=0] (7.14,-3.43) -- (0,0) -- (7.14,3.43) -- cycle    ;
\draw  [color={rgb, 255:red, 74; green, 144; blue, 226 }  ,draw opacity=1 ][line width=1.5]  (415,340) .. controls (415,323.43) and (428.43,310) .. (445,310) .. controls (461.57,310) and (475,323.43) .. (475,340) .. controls (475,356.57) and (461.57,370) .. (445,370) .. controls (428.43,370) and (415,356.57) .. (415,340) -- cycle ;
\draw  [dash pattern={on 0.84pt off 2.51pt}]  (445,340) -- (446,407) ;
\draw  [dash pattern={on 0.84pt off 2.51pt}]  (445,341) -- (392,341) ;
\draw  [color={rgb, 255:red, 74; green, 144; blue, 226 }  ,draw opacity=1 ][line width=1.5]  (325.5,282) .. controls (325.5,269.57) and (335.57,259.5) .. (348,259.5) .. controls (360.43,259.5) and (370.5,269.57) .. (370.5,282) .. controls (370.5,294.43) and (360.43,304.5) .. (348,304.5) .. controls (335.57,304.5) and (325.5,294.43) .. (325.5,282) -- cycle ;
\draw  [dash pattern={on 0.84pt off 2.51pt}]  (348,283) -- (348,410) ;
\draw  [dash pattern={on 0.84pt off 2.51pt}]  (394,282) -- (348,282) ;
\draw [color={rgb, 255:red, 65; green, 117; blue, 5 }  ,draw opacity=1 ]   (332,268) -- (364,298) ;
\draw [color={rgb, 255:red, 65; green, 117; blue, 5 }  ,draw opacity=1 ]   (425,318) -- (465,362) ;
\draw    (39,418) -- (39,231.98) ;
\draw [shift={(39,228.98)}, rotate = 90] [fill={rgb, 255:red, 0; green, 0; blue, 0 }  ][line width=0.08]  [draw opacity=0] (7.14,-3.43) -- (0,0) -- (7.14,3.43) -- cycle    ;
\draw    (26,405) -- (247,405) ;
\draw [shift={(250,405)}, rotate = 180] [fill={rgb, 255:red, 0; green, 0; blue, 0 }  ][line width=0.08]  [draw opacity=0] (7.14,-3.43) -- (0,0) -- (7.14,3.43) -- cycle    ;
\draw  [color={rgb, 255:red, 155; green, 155; blue, 155 }  ,draw opacity=0.73 ] (71.04,353.59) .. controls (56.85,321.25) and (71.18,283.36) .. (103.04,268.97) .. controls (134.89,254.57) and (172.21,269.12) .. (186.4,301.46) .. controls (200.58,333.8) and (186.25,371.68) .. (154.4,386.08) .. controls (122.54,400.48) and (85.22,385.93) .. (71.04,353.59) -- cycle ;
\draw   (200.41,294.89) .. controls (203.7,302.12) and (200.09,310.92) .. (192.34,314.54) .. controls (184.6,318.17) and (175.66,315.25) .. (172.38,308.01) .. controls (169.09,300.78) and (172.71,291.98) .. (180.45,288.36) .. controls (188.19,284.73) and (197.13,287.66) .. (200.41,294.89) -- cycle ;
\draw   (191.43,283.83) .. controls (194.32,290.18) and (191.14,297.91) .. (184.34,301.09) .. controls (177.54,304.28) and (169.69,301.71) .. (166.81,295.36) .. controls (163.92,289) and (167.09,281.27) .. (173.89,278.09) .. controls (180.69,274.91) and (188.54,277.47) .. (191.43,283.83) -- cycle ;
\draw   (203.31,309.5) .. controls (206.2,315.85) and (203.02,323.58) .. (196.22,326.77) .. controls (189.43,329.95) and (181.57,327.38) .. (178.69,321.03) .. controls (175.8,314.68) and (178.98,306.95) .. (185.78,303.76) .. controls (192.58,300.58) and (200.43,303.15) .. (203.31,309.5) -- cycle ;
\draw   (182.55,274.18) .. controls (185.17,279.96) and (182.37,286.95) .. (176.29,289.79) .. controls (170.22,292.64) and (163.16,290.26) .. (160.54,284.49) .. controls (157.92,278.71) and (160.72,271.73) .. (166.79,268.88) .. controls (172.87,266.04) and (179.92,268.41) .. (182.55,274.18) -- cycle ;
\draw   (203.36,323.07) .. controls (205.98,328.84) and (203.18,335.83) .. (197.1,338.67) .. controls (191.03,341.52) and (183.98,339.14) .. (181.35,333.37) .. controls (178.73,327.6) and (181.53,320.61) .. (187.61,317.76) .. controls (193.68,314.92) and (200.73,317.29) .. (203.36,323.07) -- cycle ;
\draw   (170.71,268.6) .. controls (172.95,273.52) and (170.83,279.35) .. (165.99,281.62) .. controls (161.14,283.89) and (155.4,281.74) .. (153.16,276.82) .. controls (150.93,271.9) and (153.05,266.07) .. (157.89,263.8) .. controls (162.74,261.53) and (168.48,263.68) .. (170.71,268.6) -- cycle ;
\draw   (199.64,335.99) .. controls (201.88,340.91) and (199.76,346.74) .. (194.91,349.01) .. controls (190.07,351.28) and (184.33,349.13) .. (182.09,344.21) .. controls (179.86,339.29) and (181.97,333.46) .. (186.82,331.19) .. controls (191.67,328.92) and (197.41,331.07) .. (199.64,335.99) -- cycle ;
\draw   (160.34,265.28) .. controls (162.14,269.25) and (160.43,273.94) .. (156.53,275.77) .. controls (152.63,277.59) and (148.01,275.86) .. (146.21,271.9) .. controls (144.41,267.94) and (146.11,263.24) .. (150.01,261.42) .. controls (153.92,259.59) and (158.54,261.32) .. (160.34,265.28) -- cycle ;
\draw   (195.06,345.89) .. controls (196.86,349.85) and (195.16,354.55) .. (191.26,356.37) .. controls (187.35,358.2) and (182.73,356.47) .. (180.93,352.51) .. controls (179.13,348.55) and (180.84,343.85) .. (184.74,342.03) .. controls (188.64,340.2) and (193.26,341.93) .. (195.06,345.89) -- cycle ;
\draw   (151.2,264.01) .. controls (152.54,266.97) and (151.27,270.48) .. (148.35,271.84) .. controls (145.43,273.21) and (141.98,271.91) .. (140.64,268.95) .. controls (139.29,265.99) and (140.56,262.49) .. (143.48,261.12) .. controls (146.4,259.75) and (149.85,261.05) .. (151.2,264.01) -- cycle ;
\draw   (190.1,354.14) .. controls (191.44,357.1) and (190.17,360.6) .. (187.25,361.97) .. controls (184.33,363.33) and (180.88,362.04) .. (179.53,359.08) .. controls (178.19,356.12) and (179.46,352.61) .. (182.38,351.25) .. controls (185.3,349.88) and (188.75,351.17) .. (190.1,354.14) -- cycle ;
\draw  [dash pattern={on 0.84pt off 2.51pt}]  (186.4,301.45) -- (185.93,402.8) ;
\draw  [dash pattern={on 0.84pt off 2.51pt}]  (186.4,301.45) -- (38.84,301.22) ;
\draw [line width=0.75]  [dash pattern={on 0.84pt off 2.51pt}]  (145.74,221.34) -- (233.82,394.12) ;
\draw [color={rgb, 255:red, 74; green, 144; blue, 226 }  ,draw opacity=1 ][line width=1.5]    (143.48,261.12) .. controls (172.25,254.33) and (194.48,283.85) .. (200.41,294.89) ;
\draw [color={rgb, 255:red, 74; green, 144; blue, 226 }  ,draw opacity=1 ][line width=1.5]    (200.41,294.89) .. controls (208.16,317.71) and (209.02,343.76) .. (187.25,361.97) ;
\draw [color={rgb, 255:red, 74; green, 144; blue, 226 }  ,draw opacity=1 ][line width=1.5]    (141.3,270.09) .. controls (158.37,281.52) and (169.68,294.27) .. (172.38,308.01) ;
\draw [color={rgb, 255:red, 74; green, 144; blue, 226 }  ,draw opacity=1 ][line width=1.5]    (172.38,308.01) .. controls (179.94,321.19) and (182.05,331.13) .. (179.08,357.74) ;
\draw  [draw opacity=0][line width=1.5]  (141.29,270.08) .. controls (139.57,267.94) and (139.37,264.91) .. (140.99,262.72) .. controls (142.51,260.67) and (145.18,259.97) .. (147.58,260.79) -- (145.92,266.48) -- cycle ; \draw  [color={rgb, 255:red, 74; green, 144; blue, 226 }  ,draw opacity=1 ][line width=1.5]  (141.29,270.08) .. controls (139.57,267.94) and (139.37,264.91) .. (140.99,262.72) .. controls (142.51,260.67) and (145.18,259.97) .. (147.58,260.79) ;  
\draw  [draw opacity=0][line width=1.5]  (190.11,359.07) .. controls (189.05,361.63) and (186.49,363.23) .. (183.82,362.78) .. controls (181.31,362.37) and (179.45,360.27) .. (179.07,357.71) -- (184.81,356.61) -- cycle ; \draw  [color={rgb, 255:red, 74; green, 144; blue, 226 }  ,draw opacity=1 ][line width=1.5]  (190.11,359.07) .. controls (189.05,361.63) and (186.49,363.23) .. (183.82,362.78) .. controls (181.31,362.37) and (179.45,360.27) .. (179.07,357.71) ;  
\draw  [dash pattern={on 0.84pt off 2.51pt}]  (186.78,217) -- (186.4,301.45) ;
\draw  [draw opacity=0] (157.43,246.37) .. controls (161.65,243.87) and (166.43,242.32) .. (171.55,241.99) .. controls (177.1,241.64) and (182.43,242.76) .. (187.23,245.06) -- (173.89,279.83) -- cycle ; \draw   (157.43,246.37) .. controls (161.65,243.87) and (166.43,242.32) .. (171.55,241.99) .. controls (177.1,241.64) and (182.43,242.76) .. (187.23,245.06) ;  
\draw  [dash pattern={on 0.84pt off 2.51pt}]  (139.75,263.88) -- (130.34,328.13) ;
\draw  [dash pattern={on 0.84pt off 2.51pt}]  (130.34,328.13) -- (181.65,362.86) ;
\draw    (141.46,335.95) .. controls (141.46,332.47) and (148.3,322.05) .. (132.05,314.24) ;

\draw (85.37,164.72) node [anchor=north west][inner sep=0.75pt]  [font=\small]  {$x_{c}$};
\draw (16.5,91.91) node [anchor=north west][inner sep=0.75pt]  [font=\small]  {$y_{c}$};
\draw (112,120.4) node [anchor=north west][inner sep=0.75pt]  [font=\small]  {$\textcolor[rgb]{0.25,0.46,0.02}{w}$};
\draw (376.37,165.72) node [anchor=north west][inner sep=0.75pt]  [font=\small]  {$x_{c}$};
\draw (305.5,90.91) node [anchor=north west][inner sep=0.75pt]  [font=\small]  {$y_{c}$};
\draw (353,48.9) node [anchor=north west][inner sep=0.75pt]  [font=\small]  {$\textcolor[rgb]{0.25,0.46,0.02}{w}\textcolor[rgb]{0.25,0.46,0.02}{_{m}}$};
\draw (407.27,83.39) node [anchor=north west][inner sep=0.75pt]  [font=\small,rotate=-1.83]  {$\alpha $};
\draw (426.5,61.91) node [anchor=north west][inner sep=0.75pt]  [font=\small]  {$\textcolor[rgb]{0.25,0.46,0.02}{w}\textcolor[rgb]{0.25,0.46,0.02}{_{M}}$};
\draw (442.37,410.72) node [anchor=north west][inner sep=0.75pt]  [font=\small]  {$x_{2}$};
\draw (374.5,328.91) node [anchor=north west][inner sep=0.75pt]  [font=\small]  {$y_{2}$};
\draw (467,365.4) node [anchor=north west][inner sep=0.75pt]  [font=\small]  {$\textcolor[rgb]{0.25,0.46,0.02}{w}\textcolor[rgb]{0.25,0.46,0.02}{_{2}}$};
\draw (396,272) node [anchor=north west][inner sep=0.75pt]  [font=\small] [align=left] {
$
y_{1}
$};
\draw (341,408) node [anchor=north west][inner sep=0.75pt]  [font=\small] [align=left] {
$
x_{1}
$};
\draw (314,245) node [anchor=north west][inner sep=0.75pt]  [font=\small] [align=left] {
$
\textcolor[rgb]{0.25,0.46,0.02}{w}\textcolor[rgb]{0.25,0.46,0.02}{_{1}}
$};
\draw (177.17,408.29) node [anchor=north west][inner sep=0.75pt]  [font=\small]  {$x_{c}$};
\draw (18.23,298.85) node [anchor=north west][inner sep=0.75pt]  [font=\small]  {$y_{c}$};
\draw (166.68,226.97) node [anchor=north west][inner sep=0.75pt]  [font=\small]  {$\alpha $};
\draw (143.16,311.11) node [anchor=north west][inner sep=0.75pt]  [font=\small]  {$\beta $};

\end{tikzpicture}

\end{center}

\caption{Gaussian shapes considered in the parametric imaging process. Top left: Gaussian circular source (circle); top right: Gaussian elliptical source (ellipse); bottom left: replication of 11 circles (loop); bottom right: double Gaussian circular source (double).}\label{fig:fig-1}
\end{figure*}
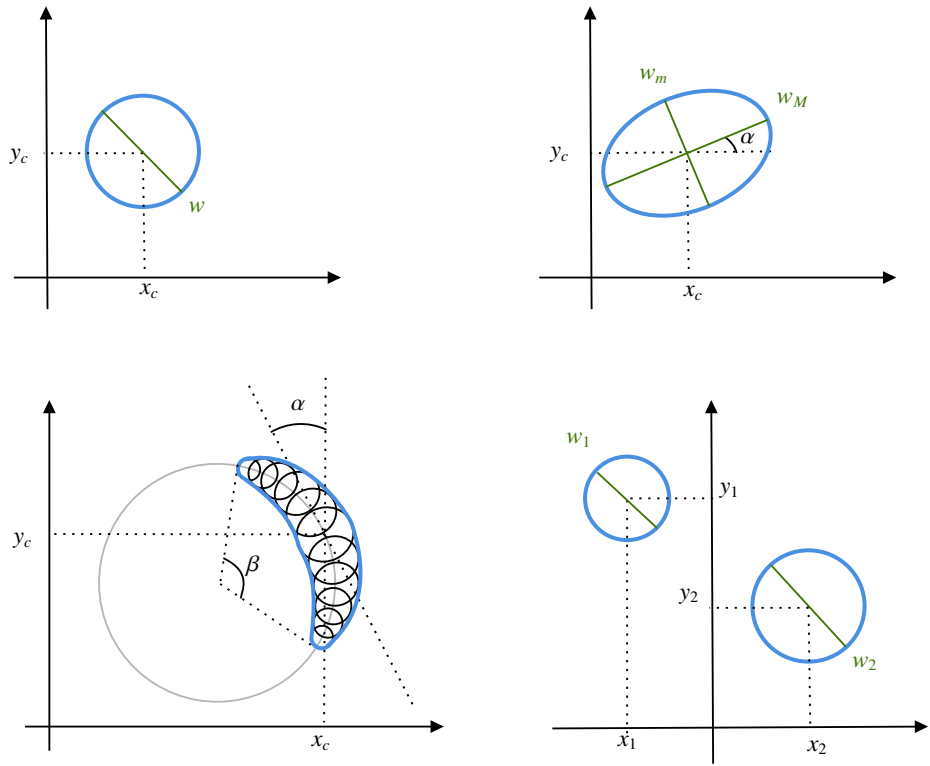

The mathematical equation describing the image formation problem for STIX is
\begin{equation}\label{b1}
\mathcal{F} \phi = {\mathbf{V}} ~,
\end{equation}
where the function $\phi = \phi(x,y)$ represents the intensity of the X-ray photon flux originating from the $(x,y)$ location on the Sun, $\mathbf{V}$ is the array containing the $N_v$ complex values of the visibilities measured by STIX and $\mathcal{F}$ is the Fourier transform defined by
\begin{equation}
(\mathcal{F}\phi)_l = \iint \phi(x,y) \exp\left( 2\pi i (x u_l + y v_l)\right) \,dx\,dy \quad l=1,\ldots, N_v ~,
\end{equation}
where $\{(u_l,v_l)\}_{l=1}^{N_v}$ is the set of angular frequencies sampled by the instrument.
Therefore, the image reconstruction problem for STIX is the linear problem of determining the photon flux from a few experimental Fourier components.
The STIX imaging problem in Eq. (\ref{b1}) and, more in general, the image reconstruction problem from hard X-ray visibilities, suffers from non-uniqueness of the solution (due to the limited $(u,v)$ coverage of the telescopes) and from ill-conditioning.
Hence, for overcoming these issues and determining a reliable solution of the imaging problem, several regularization methods have been implemented in the last decades.
We refer the reader to \cite{pianabook} for a recent review of hard X-ray imaging techniques developed so far.

In this paper we focus on forward fitting methods, whose goal is to provide estimates of the parameter values of parametric shapes used for modelling a flaring source. Physical considerations about hard X-rays emission and the experience gained during the RHESSI mission suggest that appropriate parametric shapes are those whose contour levels are shown in Figure \ref{fig:fig-1}:

\begin{itemize}

\item  A Gaussian circular source ({\em{circle}} from now on), i.e., a bi-dimensional isotropic Gaussian function described by the array of parameters $\left( x_c, y_c, \varphi, w \right)$ representing the $x$ and $y$ coordinates of the center, the total flux and the Full Width at Half Maximum (FWHM), respectively.

\item A Gaussian elliptical source ({\em{ellipse}} from now on), i.e., a bi-dimensional Gaussian function described by the array of parameters $\left( x_c, y_c, \varphi, w_M, w_m, \alpha \right)$ representing the $x$ and $y$ coordinates of the center, the total flux, the major and minor FWHM and the orientation angle, respectively.

\item A double Gaussian circular source ({\em{double}} from now on), consisting of the sum of two bi-dimensional isotropic Gaussian functions. In this configuration the parameters to optimize are the same as in the circle for each source.

\item The replication of a number of circles (in this application we considered 11 circles) with centers located along a circumference to represent a thermal loop ({\em{loop}} from now on). 
This configuration is described by the same parameters as for the ellipse with the addition of the loop angle $\beta$, i.e., the angle centered in the center of the circumference representing the curvature and subtended by the loop\footnote{\url{https://hesperia.gsfc.nasa.gov/rhessi3/software/imaging-software/vis-fwdfit/index.html}}.

\end{itemize}

We denote by ${ \boldsymbol{\theta}}$ the array of parameters characterizing each source shape. 
Then, forward-fitting STIX visibilities for parametric imaging requires the solution of the optimization problem 
\begin{equation}\label{min-prob}
\argmin_{\boldsymbol{\theta} \in \Theta}  \hskip 0.2cm \chi^2({ \boldsymbol{\theta}}), 
\end{equation}
where $\Theta$ is the parameter space and the target function $\chi^2({ \boldsymbol{\theta}})$ is defined as

\begin{equation}\label{chi2}
\chi^2({ \boldsymbol{\theta}}) = \frac{1}{N_v-N_{{ \boldsymbol{\theta}}}} \sum_{l=1}^{N_v} \frac{\left | V_l - (\mathcal{F} \phi_{\boldsymbol{\theta}})_l \right |^2}{\sigma_l^2}  \ ,
\end{equation}

where $N_{{ \boldsymbol{\theta}}}$ is the number of parameters of the source shape function $\phi_{\boldsymbol{\theta}}$ (either a circle, 
an ellipse, a double or a loop) and $\sigma_l$ is the uncertainty on the $l-$th measured visibility amplitude. 

The parametric imaging problem for STIX consists in selecting a specific shape function $\phi_{\boldsymbol{\theta}}$ as candidate source, solving problem in Eqs. (\ref{min-prob}) and (\ref{chi2}), and feeding the corresponding source shape with the solution parameters for obtaining the reconstructed image. The computational core of this procedure is the optimization scheme applied for solving Eqs. (\ref{min-prob}) and (\ref{chi2}).
In the Solar SoftWare (SSW) repository the VIS$\_$FWDFIT routine utilizes the simplex scheme (AMOEBA) and has been applied for the parametric imaging of all RHESSI data sets published so far and for the creation of the RHESSI Image Archive\footnote{\url{https://hesperia.gsfc.nasa.gov/rhessi3/mission-archive/index.html}}. The effectiveness of this approach significantly decreases when the number of the available data becomes smaller, as is the case of STIX measurements. Therefore, in this framework, new optimization approaches must be introduced, which are able to avoid local minima by means of a more effective exploration of the parameter space.

\section{Particle Swarm Optimization}

Particle Swarm Optimization (PSO) \citep{eberhart1995particle,Qasem} realizes optimization by mimicking swarm intelligence. The starting point of swarm intelligence is a random initializiation of a set of points within the parameter space $\Theta$, which gives rise to a \emph{swarm of particles} or \emph{birds}.
Then, the position of each bird is iteratively modified  so that the swarm accumulates around the point of $\Theta$ corresponding to the minimum of the objective function (the $\chi^2$ function in the parametric STIX imaging).
Specifically, at each iteration, the location of each particle is updated based on its velocity at the previous iteration, the best position visited by the particle since the beginning of the iterative process and the global best position visited by the whole swarm.
For details on the implementation of this procedure we refer the reader to Section 3 in \cite{2021A&A...656A..25M}.

The IDL PSO routine we have implemented and included in SSW is based on the implementation introduced by \cite{Coello}. Differently than for VIS$\_$FWDFIT, VIS$\_$FWDFIT$\_$PSO has been implemented in such a way that the user can decide which parameters have to be optimized while keeping the other ones fixed. 
In both VIS$\_$FWDFIT and VIS$\_$FWDFIT$\_$PSO the uncertainty on the parameters is estimated by
generating a confidence strip on each parameter value: several realizations of the input data are computed by randomly perturbing the experimental set of visibility 
with Gaussian noise whose standard deviation is set equal to the errors on the measurements; for each realization the optimization method is applied; and, finally, the standard deviation of each optimized source parameter is computed.

\section{Numerical and experimental results}

We have compared the performances of VIS$\_$FWDFIT$\_$PSO with respect to the ones of VIS$\_$FWDFIT in the case of three experiments. First, we verified that PSO parameter estimates are comparable with the ones provided by AMOEBA in the case of RHESSI visibilities, when VIS$\_$FWDFIT showed notable reliability and robustness. Then, we generated synthetic STIX visibilities associated with four source configurations mimicking typical non-thermal emissions and we applied VIS$\_$FWDFIT$\_$PSO and VIS$\_$FWDFIT to these four simulated data sets. Finally, we compared the robustness of the two optimization schemes with respect to different choices of the map center in the case of STIX experimental visibilities observed during the SOL2021-08-26T23:20 event.

\subsection{RHESSI visibilities}

\begin{center}
    \begin{figure}[]
        \centering
          \includegraphics[height=\textwidth, angle=90]{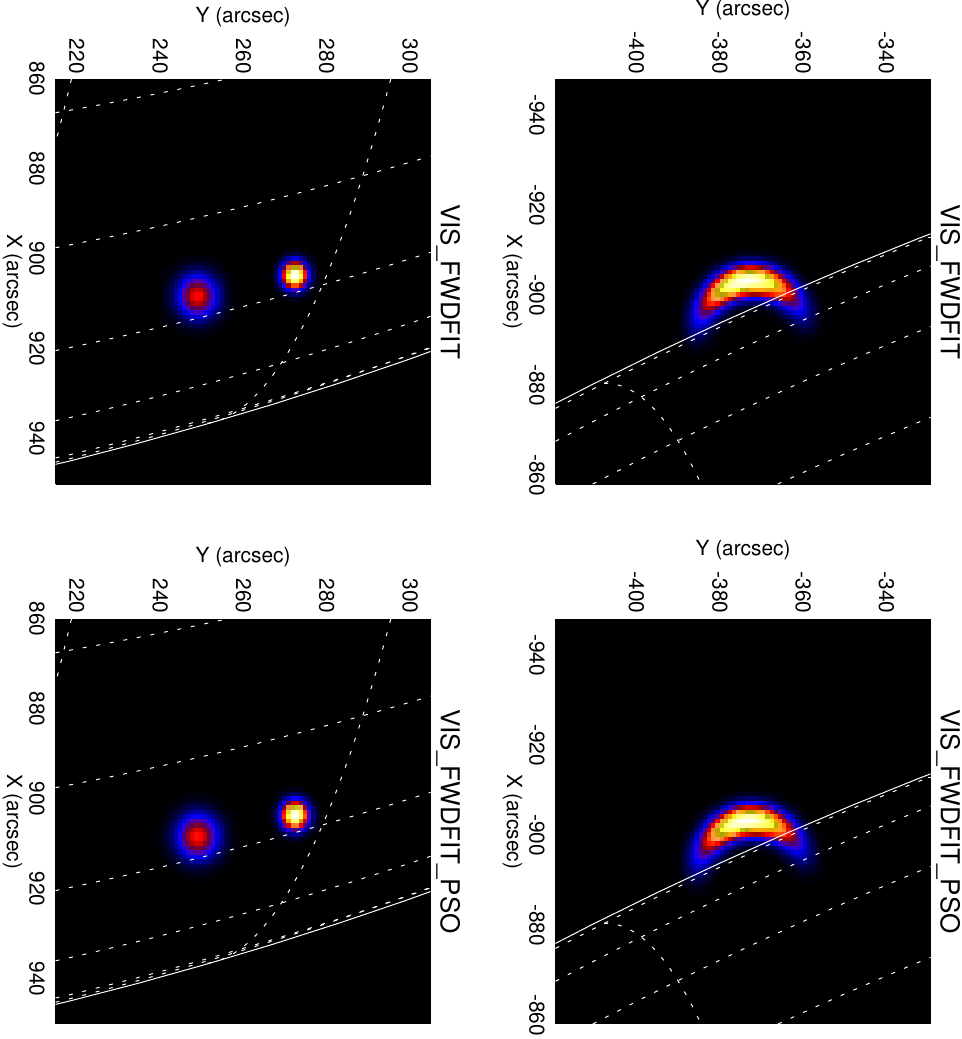}
        \caption{Forward fitting RHESSI visibilities with AMOEBA and PSO. Top row: reconstructions of the thermal component (6 -- 12 keV) of the February 13 2002 event 12:29:40 -- 12:31:22 UT provided by VIS$\_$FWDFIT (left) and VIS$\_$FWDFIT$\_$PSO (right). Bottom row: reconstruction of the non-thermal component (22 -- 50 keV) of the February 20 2002 event 11:06:05 -- 11:07:42 UT. }        \label{fig:RHESSI}
    \end{figure}
\end{center}

On February 13 2002, in the time range between 12:29:40 UT and 12:31:22 UT a flaring emission of class C1.3 showed a loop behavior in the thermal energy range between 6 and 12 keV. RHESSI well observed both such emission and also, on the 20th of the same month, a C7.5 limb flare that, in the time range 11:06:05 -- 11:07:42 UT, was characterized by a clearly visible  non-thermal component in the energy range 22-50 keV. Figure \ref{fig:RHESSI} shows the reconstructions of these two components provided by VIS$\_$FWDFIT and VIS$\_$FWDFIT$\_$PSO, while Table \ref{RHESSI-thermal} and Table \ref{RHESSI-non-thermal} contain the estimated values for the imaging parameters.

\begin{table}[]
\centering
\resizebox{\columnwidth}{!}{\begin{tabular}{cccccc}
\hline\hline
energy (keV) & method & shape & flux (counts/s/keV) & orientation (deg) & curvature (deg)  \\
\hline
6-12 & VIS\_FWDFIT & loop & 323.7 $\pm$ 12.9 & 6.1 $\pm$ 5.3 & -78.5 $\pm$ 27.6 \\
6-12 & VIS\_FWDFIT\_PSO & loop & 325.0 $\pm$ 12.4 & 6.1 $\pm$ 5.2 & -77.8 $\pm$ 29.4 \\
\hline
\end{tabular}}
\caption{February, 13 2002 flare observed by RHESSI in the energy range 6--12 keV and in the time interval 12:29:40 -- 12:31:22 UT. The parameter values are provided by VIS$\_$FWDFIT and VIS$\_$FWDFIT$\_$PSO when the chosen shape is loop.}
\label{RHESSI-thermal}
\end{table}
    
\begin{table}[]
    \centering
    \resizebox{\columnwidth}{!}{\begin{tabular}{cccccc}
    \hline\hline 
energy (keV) & method & shape & component & flux (counts/s/keV) & FWHM (arcsec) \\    
\hline
22-50 & VIS\_FWDFIT & double & 1 & 11.0 $\pm$ 0.5 & 4.6 $\pm$ 0.6 \\
22-50 & VIS\_FWDFIT\_PSO & double & 1 & 11.4 $\pm$ 0.4 & 4.8 $\pm$ 0.5 \\
22-50 & VIS\_FWDFIT & double & 2 & 13.1 $\pm$ 0.5 & 7.6 $\pm$ 0.6 \\
22-50 & VIS\_FWDFIT\_PSO & double & 2 & 13.7 $\pm$ 0.6 & 7.9 $\pm$ 0.7 \\
\hline
\end{tabular}}
        \caption{February, 20 2002 flare observed by RHESSI in the energy range 22--50 keV and in the time interval 11:06:05 -- 11:07:42 UT. The parameter values are provided by VIS$\_$FWDFIT and VIS$\_$FWDFIT$\_$PSO when the chosen shape is double.}
        \label{RHESSI-non-thermal}
\end{table}

\subsection{PSO and AMOEBA for STIX synthetic visibilities}

Synthetic STIX visibilities can be generated utilizing the STIX simulation software. Once a specific flare configuration is selected (e.g. circle, ellipse, double, loop), a Monte Carlo approach is used to reproduce the trajectory of the photons counts emitted from the source and measured by the detector pixels. The visibility and related uncertainty are then computed from the simulated counts measurements.

\begin{figure}
    \centering
      \includegraphics[height=\textwidth, angle=90]{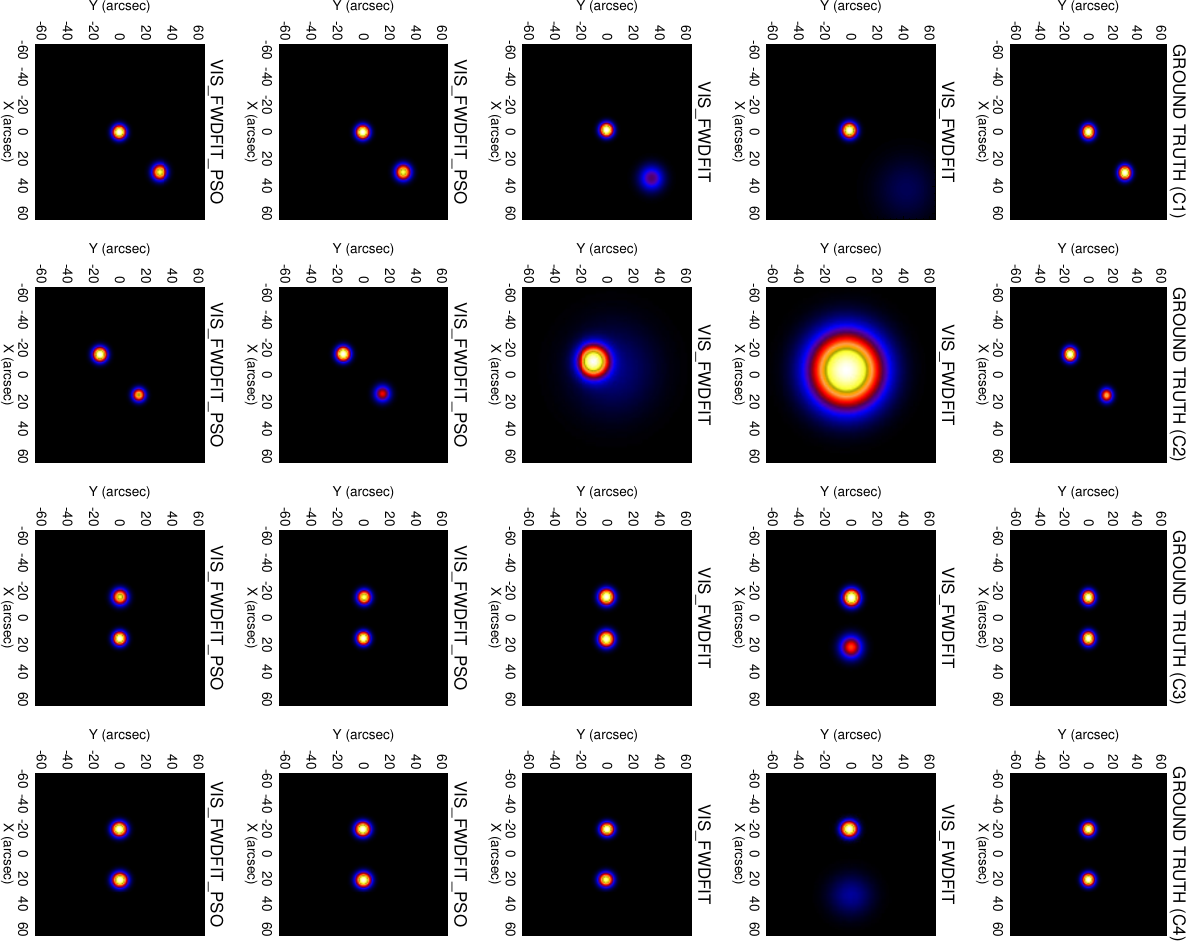}
    \caption{First row: simulated configurations mimicking four non-thermal emissions (the values of the configuration parameters are given in Table \ref{AMOEBA and PSO}). Second and third rows: the two most frequent behaviours of VIS\_FWDFIT reconstructions. Fourth and fifth rows: the two most frequent behaviours of VIS\_FWDFIT\_PSO reconstructions.}
    \label{synthetic}
\end{figure}

We have simulated four configurations mimicking four non-thermal emissions, in all cases represented by two foot-points. Specifically, from left to right in the first row of Figure \ref{synthetic}, the first panel (configuration C1) contains two foot-points that have the same flux of 5000 photons s$^{-1}$ keV$^{-1}$, the same FWHM (8 arcsec), and are located at (0,0) and (30,30) arcsec, respectively. In the second panel (configuration C2), the source on the top right has a reduced flux equal to 3000 photons s$^{-1}$ keV$^{-1}$, the FHWM is the same, and the positions are (-15,-15) and (15,15) arcsec, respectively. In the third and fourth panels (configurations C3 and C4, respectively), the two foot-points with the same flux are now placed at distances of 30 and 40 arcsec, respectively. We have generated 25 random realizations of the visibility sets corresponding to the four configurations and applied VIS$\_$FWDFIT and VIS$\_$FWDFIT$\_$PSO to them. As far as VIS$\_$FWDFIT is concerned, most of the times the algorithm provides reconstructions that behave as in the second and third rows of Figure \ref{synthetic}; on the other hand, VIS$\_$FWDFIT$\_$PSO behaves most of the times as in the examples included in the fourth and fifth rows of the figure, respectively. Finally, Table \ref{AMOEBA and PSO} contains the parameters estimated by the two methods for the four configurations. 

\begin{center}
\begin{table}
\resizebox{\columnwidth}{!}{\begin{tabular}{ccccccccccc}
\hline\hline
& & \multicolumn{4}{ c }{First source} & & \multicolumn{4}{ c }{Second source} \\
\cline{3-6}
\cline{8-11}
& & flux & FWHM & x & y & & flux & FWHM & x & y\\
&& (counts/s/keV) & (arcsec) & (arcsec) & (arcsec) & & (counts/s/keV) & (arcsec) & (arcsec) & (arcsec) \\
\hline
 & Simulated & 5000.0 & 8.0 & 0.0 & 0.0 & &
5000.0 & 8.0 & 30.0 & 30.0 \\
C1 & VIS\_FWDFIT & 4850.8 $\pm$ 458.9 & 8.0 $\pm$ 2.0 & -1.0 $\pm$ 0.4 & -0.2 $\pm$ 0.6 &
& 5139.2 $\pm$ 497.6 & 20.7 $\pm$ 11.1 & 35.1 $\pm$ 5.4 & 35.1 $\pm$ 5.4 \\
& VIS\_FWDFIT\_PSO & 4778.3 $\pm$ 85.8 & 8.5 $\pm$ 0.7 & -0.1 $\pm$ 0.2 & 0.1 $\pm$ 0.2 &
   & 4838.1 $\pm$ 58.0 & 8.7 $\pm$ 0.4 & 30.1 $\pm$ 0.2 & 30.1 $\pm$ 0.2 \\
\hline
& Simulated & 5000.0 & 8.0 & -15.0 & -15.0 & & 3000.0 & 8.0 & 15.0 & 15.0 \\
C2 & VIS\_FWDFIT & 4385.4 $\pm$ 2505.7 & 36.7 $\pm$ 10.6 & -4.5 $\pm$ 3.3 & -4.5 $\pm$ 3.3 &
& 3664.6 $\pm$ 2466.5 & 39.7 $\pm$ 10.9 & -2.00 $\pm$ 4.6 & -1.3 $\pm$ 4.9 \\
& VIS\_FWDFIT\_PSO & 4877.9 $\pm$ 84.1 & 8.4 $\pm$ 0.6 & -14.9 $\pm$ 0.2 & -15.0 $\pm$ 0.2 &
& 2840.3 $\pm$ 85.4 & 8.3 $\pm$ 0.9 & 14.9 $\pm$ 0.5 & 14.9 $\pm$ 0.4 \\
\hline
& Simulated & 5000.0 & 8.0 & -15.0 & 0.0 & & 5000.0 & 8.0 & 15.0 & 0.0 \\
C3 & VIS\_FWDFIT & 5091.9 $\pm$ 257.1 & 9.6 $\pm$ 1.2 & -14.9 $\pm$ 0.4 & -0.1 $\pm$ 0.4 &  
& 4691.3 $\pm$ 162.1 & 11.7 $\pm$ 3.8 & 18.4 $\pm$ 3.6 & 0.1 $\pm$ 0.3 \\
& VIS\_FWDFIT\_PSO & 4817.1 $\pm$ 156.7 & 8.5 $\pm$ 0.6 & -15.1 $\pm$ 0.2 & 0.1 $\pm$ 0.3 & &4807.3 $\pm$ 133.4 & 8.4 $\pm$ 0.6 & 15.0 $\pm$ 0.3 & -0.1 $\pm$ 0.2 \\
\hline
& Simulated & 5000.0 & 8.0 & -20.0 & 0.0 & &5000.0 & 8.0 & 20.0 & 0.0 \\
C4 & VIS\_FWDFIT & 5224.1 $\pm$ 449.1 & 9.1 $\pm$ 0.8 & -19.9 $\pm$ 0.3 & -1.0 $\pm$ 0.9& 
& 4782.5 $\pm$ 111.8 & 16.1 $\pm$ 7.7 & 25.9 $\pm$ 6.1 & 0.2 $\pm$ 0.4 \\
   & VIS\_FWDFIT\_PSO & 4834.0 $\pm$ 87.4 & 8.5 $\pm$ 0.5 & -20.0 $\pm$ 0.2 & -0.0 $\pm$ 0.3 &
   & 4819.4 $\pm$ 119.9 & 8.5 $\pm$ 0.5 & 20.0 $\pm$ 0.3 & -0.1 $\pm$ 0.2 \\
\hline
\end{tabular}}
\caption{Average values and standard deviations of the imaging parameters estimated by VIS$\_$FWDFIT and VIS$\_$FWDFIT$\_$PSO for the reconstruction of the four synthetic configurations in the first row of Figure \ref{synthetic}. For each configuration, the two algorithms have been applied to 25 random realizations of the STIX synthetic visibilities.}\label{AMOEBA and PSO}
\end{table}
\end{center}



\subsection{PSO and AMOEBA for STIX experimental visibilities}

\begin{figure}
    \centering
    \includegraphics[width=0.8\textwidth]{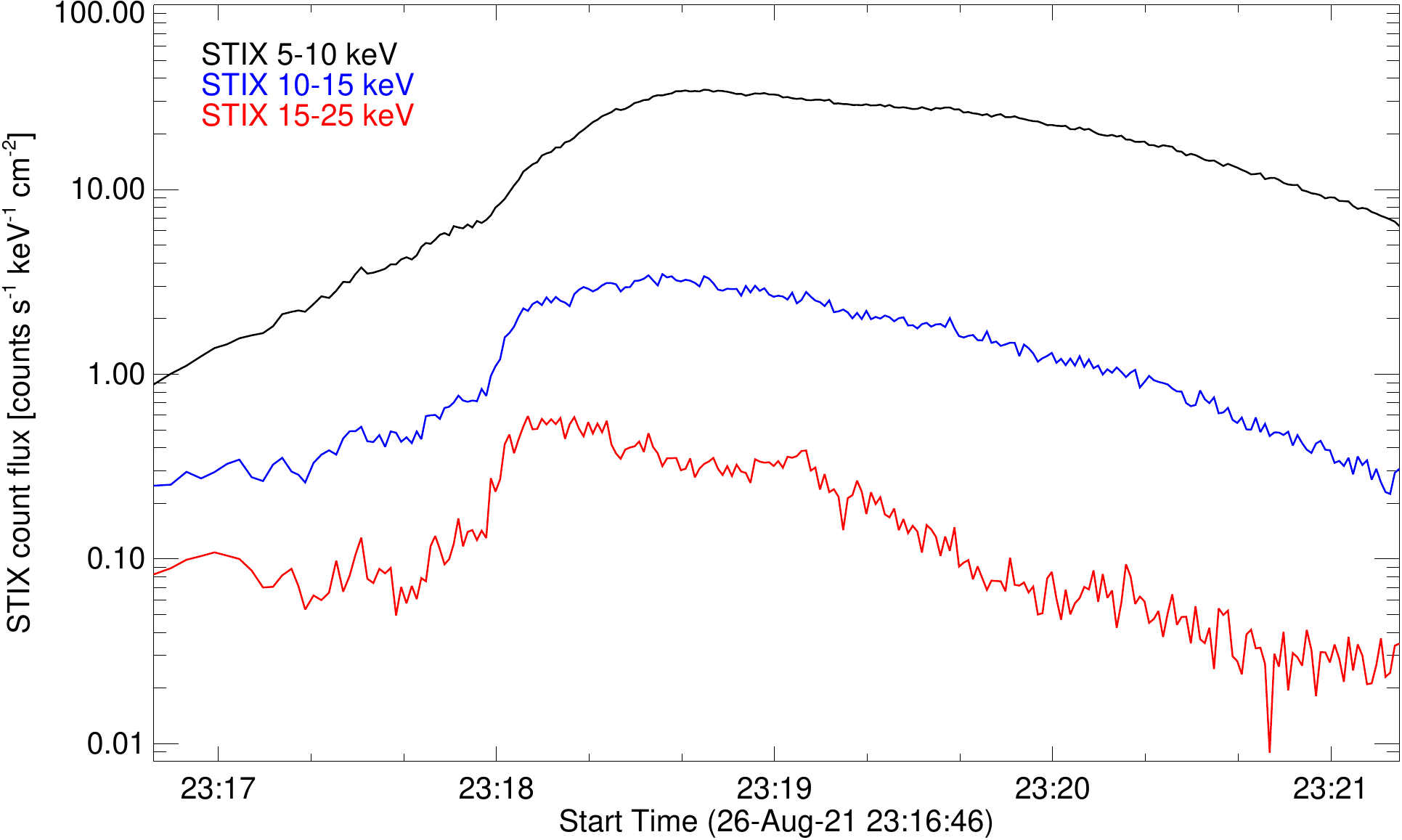}
    \caption{Light curves for the August 26, 2021 event in the time range from 23:16:46 UT to 23:21:14 UT.}
    \label{lightcurves}
\end{figure}

\begin{figure}
    \centering
    \includegraphics[width=13.cm, angle=90]{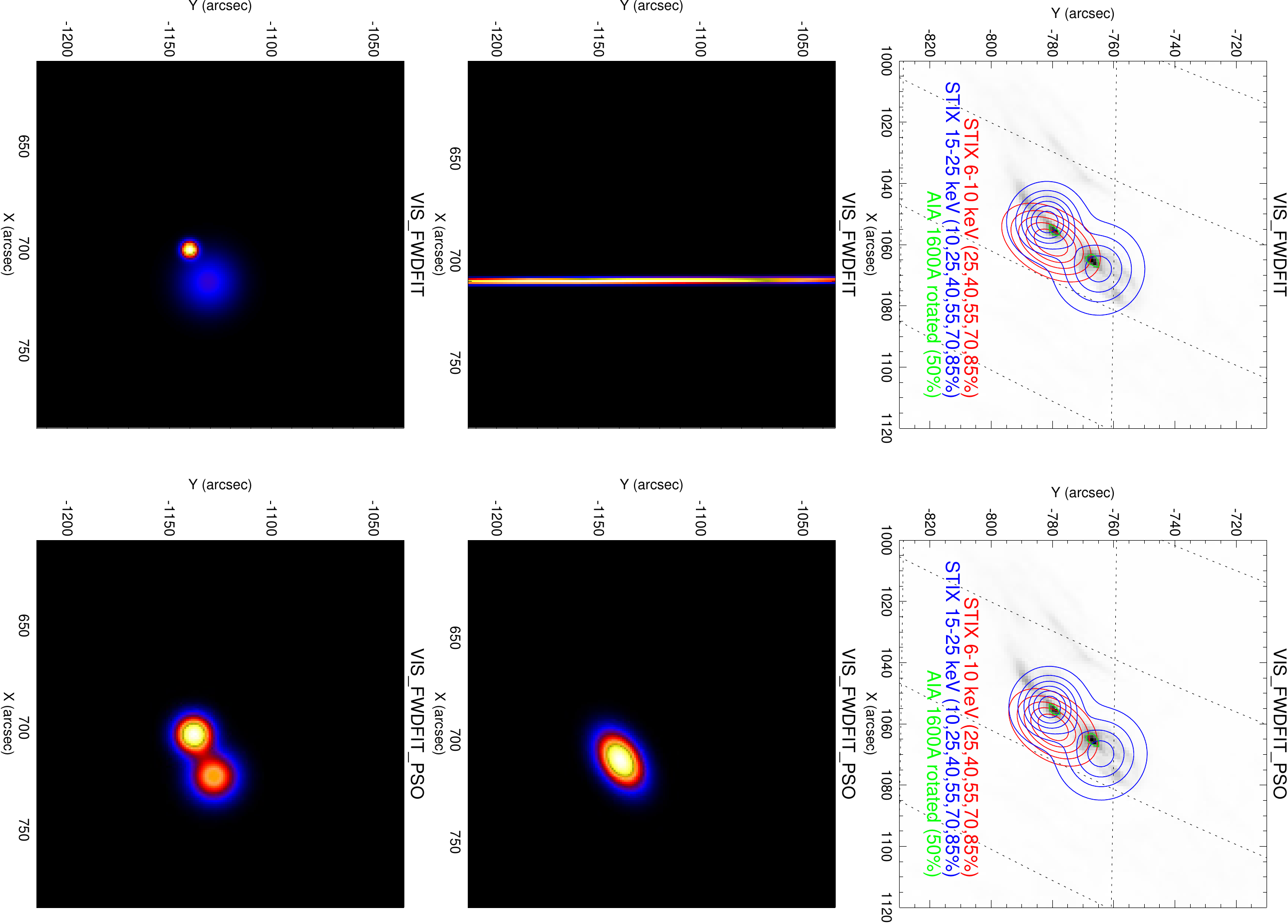}
    \caption{Parametric images of the August 26, 2021 event obtained by applying VIS\_FWDFIT (left column) and VIS\_FWDFIT\_PSO (right columns) on STIX visibilities. First row: level contour of the thermal (red) and non-thermal (blue) components superimposed onto the rotated AIA map (the 50\% contour level of the AIA image is plotted in green). Second row: parametric images of the thermal component with the map-center shifted of $ |\Delta x| = |\Delta y |= 10$ arcsec. Third row: same as the second row but in the case of the non-thermal component.}
    \label{STIX-experimental}
\end{figure}

On August, 26 2021 STIX observed a GOES C3 class flare close to the limb,
whose lightcurves are represented in Figure \ref{lightcurves}.
We have considered the data set corresponding to the time range between 23:18:00 and 23:20:00 and separately studied the thermal emission in the energy range between 6 and 10 keV and the non-thermal emission in the energy range between 15 and 25 keV. Further, for this experiment we did not use the 6 visibilities measured by the sub-collimators with finest resolution, since they are not yet fully calibrated. We have first computed the discretized inverse Fourier transform of these visibilities by applying the back projection algorithm and identified the location of the maximum of the reconstructed 'dirty maps' as reference map centers. We applied VIS$\_$FWDFIT and VIS$\_$FWDFIT$\_$PSO to the visibility set corresponding to the thermal emission using the reference map centers and obtained the reconstructions represented in the first row of Figure \ref{STIX-experimental}. Specifically, in these two panels, the level curves of the maps provided by VIS$\_$FWDFIT and VIS$\_$FWDFIT$\_$PSO are manually superimposed to the EUV maps measured by SDO/AIA in the same time interval. 
Following the procedure described in \citet{2021A&A...656A...4B}, the AIA maps have been reprojected to make them appear as they would be seen from Solar Orbiter.
Then, we again applied the two routines, but this time with the map center shifted of $\abs{\Delta x}= \abs{\Delta y} =$ 10 arcsec. The results of this second experiment are provided in the two bottom rows of the same figure. We have finally repeated this experiment several times, for several values of the shift $\Delta x$ in both directions, while keeping $\Delta y$ fixed at $\Delta y=0$. In Table \ref{fwhm} 
we have reported the values of the FWHM for each reconstruction in the non-thermal regime. We have highlighted in red all cases where the relative variation of the FWHM is bigger than 10\% with respect to the optimal case $\Delta x=$0.

\subsection{Analysis of the computational burden}

As a final validation step, we have analyzed the differences in the computational burden for the two schemes in the four cases involving the experimental RHESSI and STIX visibilities considered in this study. 
Specifically, Table \ref{time} compares the computational costs employed by VIS$\_$FWDFIT and VIS$\_$FWDFIT$\_$PSO both in the setting when the algorithms compute the uncertainties on the estimated parameters and in the setting when just one realization of the visibility set is considered.
Tests have been performed on an Apple M1 processor.

\begin{table}
\centering
\begin{tabular}{ ccccc }
\hline
\hline
Shift (arcsec)   & Source   & \multicolumn{3}{c}{FWHM (arcsec)} \\
\cline{3-5}
& &  VIS\_FWDFIT    &   & VIS\_FWDFIT\_PSO \\
\hline

\multirow{2}{6em}{$\Delta x = +20 $} & First source  & \cellcolor{red!65}22.9 $\pm$ 1.2 & &  \cellcolor{green!40}14.4 $\pm$ 1.6 \\
& Second source  & \cellcolor{red!65}12.4 $\pm$ 44.9 & &  \cellcolor{green!40}18.4 $\pm$ 2.3\\ 
\hline

\multirow{2}{ 6em}{$\Delta x = +15 $} & First source  & \cellcolor{red!65}17.8 $\pm$ 1.4& &  \cellcolor{green!40}14.4 $\pm$ 1.7\\
& Second source  & \cellcolor{red!65}16.6 $\pm$ 7.6 & &  \cellcolor{green!40}18.4 $\pm$ 1.9 \\
\hline

\multirow{2}{ 6em}{$\Delta x = +10 $} & First source  & \cellcolor{green!40}13.4 $\pm$ 1.1 & & \cellcolor{green!40}14.4 $\pm$ 1.8\\
& Second source  & \cellcolor{green!40}19.2 $\pm$ 3.4 & &  \cellcolor{green!40}18.4 $\pm$ 1.7 \\
\hline

\multirow{2}{ 6em}{$\Delta x = +5 $} & First source  & \cellcolor{red!65}11.9 $\pm$ 2.0 & &  \cellcolor{green!40}14.4 $\pm$ 1.8 \\
& Second source  & \cellcolor{red!65}22.0 $\pm$ 1.9 & &  \cellcolor{green!40}18.4 $\pm$ 2.3\\
\hline

\multirow{2}{ 6em}{$\Delta x = 0 $} & First source  & \cellcolor{green!40}14.4 $\pm$ 1.9 & & \cellcolor{green!40}14.4 $\pm$ 1.9 \\
& Second source  & \cellcolor{green!40}18.4 $\pm$ 1.7 & &  \cellcolor{green!40}18.4 $\pm$ 1.9\\
\hline

\multirow{2}{ 6em}{$\Delta x = -5 $} & First source  & \cellcolor{green!40}14.0 $\pm$ 4.0 & & \cellcolor{green!40}14.4 $\pm$ 2.2 \\
& Second source  & \cellcolor{green!40}18.7 $\pm$ 2.5 & & \cellcolor{green!40}18.4 $\pm$ 2.2 \\
\hline

\multirow{2}{ 6em}{$\Delta x = -10 $} & First source  & \cellcolor{red!65}9.7 $\pm$ 0.0 & & \cellcolor{green!40}14.5 $\pm$ 2.6\\
& Second source  & \cellcolor{red!65}22.0 $\pm$ 2.6& &  \cellcolor{green!40}18.4 $\pm$ 2.3\\
\hline

\multirow{2}{ 6em}{$\Delta x = -15 $} & First source & \cellcolor{red!65}0.52 $\pm$ 7.1 & &\cellcolor{green!40}14.4 $\pm$ 2.2\\
& Second source  & \cellcolor{red!65}25.4 $\pm$ 9.3 & &  \cellcolor{green!40}18.4 $\pm$ 2.2\\
\hline

\multirow{2}{ 6em}{$\Delta x = -20 $} & First source source & \cellcolor{red!65}0.52 $\pm$ 10.6 & &  \cellcolor{green!40}14.5 $\pm$ 1.8\\
& Second source & \cellcolor{red!65}27.6 $\pm$ 3.4 & & \cellcolor{green!40}18.4 $\pm$ 2.2 \\
\hline
\end{tabular}
\caption{FWHM associated with the parametric images of the August 26, 2021 event obtained by applying VIS\_FWDFIT and VIS\_FWDFIT\_PSO to STIX visibilities for different map center values. Each entry in the third and fourth column is the estimated FWHM value when the map-center is shifted of $\Delta x$ arcsec while keeping $\Delta y=0$. First source and second source refer to the ones at the bottom left and top right in Figure \ref{STIX-experimental}, respectively. The red (green) color points out the reconstructions for which the relative variation of the FWHM is bigger (smaller) than 10\% with respect to the optimal case $\Delta x=0$.}\label{fwhm}
\end{table}

\begin{table}[h]
\centering
\begin{tabular}{ ccccc}
\hline\hline
 Event   & Method   &  Shape    &   \multicolumn{2}{c}{Computation time (sec.)} \\
\hline
&         &         & \multicolumn{2}{c}{Uncertainty} \\
&         &         &   \ on           &  \ off \\
          \hline
13--Feb--2002 &VIS\_FWDFIT & \multirow{2}{4em}{Loop}   &  1.66 & 0.10 \\
& VIS\_FWDFIT\_PSO &              & 62.1  & 3.00  \\
\hline
20--Feb--2002 & VIS\_FWDFIT  &    \multirow{2}{4em}{Double} & 0.42   &  0.04 \\
             & VIS\_FWDFIT\_PSO &       &  37.6  &  5.20 \\
\hline
26--Aug--2021 &VIS\_FWDFIT &    \multirow{2}{4em}{Ellipse} & 0.15   & 0.07 \\
& VIS\_FWDFIT\_PSO &              &  2.54  &  0.18 \\
\hline
26--Aug--2021 &VIS\_FWDFIT &    \multirow{2}{4em}{Double} &  0.55 &  0.10  \\
& VIS\_FWDFIT\_PSO &              &  21.8 &  3.60  \\
\hline
\end{tabular}
\caption{
Computational time of VIS\_FWDFIT and VIS\_FWDFIT\_PSO in the case of the four experimental data sets considered in the paper.
Each reconstruction is performed 10 times and the average computational time is reported in the table.
The labels ``on'' and ``off'' indicate whether the uncertainty on the retrieved parameters is estimated or not, respectively.
}\label{time}
\end{table}



\section{Comments and conclusions}

The image reconstruction problem when data are hard X-ray visibilities is intrinsically linear and numerically unstable. Image reconstruction methods relying on regularization theory for ill-posed linear inverse problems have the advantage of providing reconstructions without any a priori assumption on the source morphology and typically via algorithms characterized by a relatively low computational burden. However, these linear approaches do not explicitly provide estimates of the imaging parameters. Parametric imaging relies on prior assumptions on the shape of the flaring source and often involves sophisticated optimization schemes able to deal with non-convex functionals. Their main advantage is that they automatically provides estimates on the imaging parameters together with estimates on the corresponding statistical uncertainty.

This paper proved that the AMOEBA, i.e. the optimization scheme which is at the core of the SSW VIS$\_$FWDFIT routine, and which has shown a notable reliability for the forward fitting of RHESSI visibilities, is significantly less effective in the STIX framework, where the number of visibility is significantly reduced and, coherently, the sampling of the spatial frequency plane is significantly more sparse. For this reason we have implemented and included in SSW a new forward fitting routine, namely VIS$\_$FWDFIT$\_$PSO, in which the optimization step is performed by means of a biology-inspired algorithm, namely Particle Swarm Optimization (PSO). This new routine has the same good performances with respect to VIS$\_$FWDFIT in the case of RHESSI data, but is significantly more reliable in the case of STIX visibilities. Specifically, the new routine is more robust with respect to even slight modifications of the map center and is able to provide satisfactory results in the case of the reconstruction of source configurations associated to the non-thermal regime, when the foot points are characterized by different intensities. The price to pay for this increased reliability is a heavier computational burden, which is more than one order of magnitude higher when the setting requires the computation of the uncertainties on the parameter estimates. However, this extra time is still very well within reasonable time frame for data analysis in this context.

\begin{acknowledgements}
{\em{Solar Orbiter}} is a space mission of international collaboration between ESA and NASA, operated by ESA. The STIX instrument is an international collaboration between Switzerland, Poland, France, Czech Republic, Germany, Austria, Ireland, and Italy. AFB and SK are supported by the Swiss National Science Foundation Grant 200021L\_189180 and the grant 'Activités Nationales
Complémentaires dans le domaine spatial' REF-1131-61001 for STIX. PM, EP, FB and MP acknowledge the financial contribution from the agreement ASI-INAF n.2018-16-HH.0. 
SG acknowledges the financial support from the "Accordo ASI/INAF Solar Orbiter: Supporto scientifico per la realizzazione degli strumenti Metis, SWA/DPU e STIX nelle Fasi D-E".
\end{acknowledgements}


\bibliographystyle{aa}
\bibliography{bib_stix}

\end{document}